\documentclass[final,3p,11pt]{elsarticle}

\usepackage[colorlinks=true, allcolors=blue]{hyperref}

\usepackage{fancyhdr,graphicx,amsmath,amssymb,amsthm, nccmath}
\usepackage{multirow} 
\usepackage{multicol}
\usepackage{dashrule}
\usepackage{makecell}
\usepackage{diagbox}
\usepackage{aligned-overset}

\usepackage{lipsum}
\usepackage{bm}
\usepackage{float}
\usepackage{mathrsfs}
\usepackage{enumitem}
\usepackage{booktabs} 

\usepackage{longtable}
\usepackage{bm}
\usepackage[ruled,vlined]{algorithm2e}
\SetKwInput{Parameters}{Parameters}
\SetKwInput{Initialize}{Initialization}
\usepackage{algpseudocode}

\usepackage{xcolor}
\usepackage{ulem}
\usepackage{subcaption}
\usepackage{mdframed}
\usepackage{caption}
\usepackage[T1]{fontenc}
\usepackage[utf8]{inputenc}

\usepackage{verbatim} 
\usepackage{enumitem}

\newcommand{\dashrule}[1][black]{%
  \color{#1}\rule[\dimexpr.5ex-.2pt]{4pt}{.4pt}\xleaders\hbox{\rule{4pt}{0pt}\rule[\dimexpr.5ex-.2pt]{4pt}{.4pt}}\hfill\kern0pt%
}

\usepackage{xspace}
\newcommand{\evclust}{\textsf{evclust}\xspace}

\makeatletter
\def\ps@pprintTitle{%
 \let\@oddhead\@empty
 \let\@evenhead\@empty
 \let\@oddfoot\@empty
 \let\@evenfoot\@empty
}
\makeatother

\journal{}

\begin{document}
\begin{frontmatter}

\title{\evclust: Python library for evidential clustering}

\author[1]{Armel Soubeiga}
\ead{armel.soubeiga@uca.fr}
\author[1]{Violaine Antoine}
\ead{violaine.antoine@uca.fr}

\affiliation[1]{organization={University of Clermont Auvergne, CNRS, Mines of Saint-Etienne, Clermont-Auvergne-INP, LIMOS, 63000 Clermont-Fd}, city={Clermont-Ferrand},country={France}}

\begin{abstract}
A recent developing trend in clustering is the advancement of algorithms that not only identify clusters within data, but also express and capture the uncertainty of cluster membership. Evidential clustering addresses this by using the Dempster-Shafer theory of belief functions, a framework designed to manage and represent uncertainty. This approach results in a credal partition, a structured set of mass functions that quantify the uncertain assignment of each object to potential groups. The Python framework \evclust, presented in this paper, offers a suite of efficient evidence clustering algorithms as well as tools for visualizing, evaluating and analyzing credal partitions.
\end{abstract}


\end{frontmatter}

\section{Introduction}

Clustering is a fundamental task in data analysis, aimed at grouping objects into clusters based on similarity. Traditional methods such as the k-means algorithm \citep{ev006} have dominated this field for decades, providing a simple and efficient approach to partitioning data. However, hard clustering techniques like k-means assign each object to a single cluster, which can be limiting in applications where uncertainty or overlap between clusters must be accounted. \\

Soft clustering methods address this limitation by allowing partial memberships to multiple clusters. Fuzzy clustering, for example, computes degrees of membership for each object in all clusters, as implemented in Python libraries such as \texttt{scikit-fuzzy} \citep{ev022}. Possibilistic clustering \citep{ev020}, on the other hand, relaxes the sum constraint on membership degrees, allowing better handling of outliers. Rough clustering approaches \citep{ev021} define lower and upper approximations for clusters, providing a nuanced view of cluster membership. Building on these approaches, evidential clustering is a more general framework that quantifies uncertainty using the Dempster-Shafer theory of belief functions \citep{ev013}. Instead of assigning simple degrees of membership, it uses mass functions that distribute belief over subsets of clusters. This richer representation allows evidential clustering to capture complex uncertainty patterns, such as partial ignorance, where an object may belong to multiple clusters without a clear preference \citep{ev013}. \\

While R has tools like \evclust \citep{ev014} for evidential clustering, Python lacked a comprehensive library for this purpose. To fill this gap, we introduce \evclust, a Python library that implements state-of-the-art evidential clustering algorithms. \evclust provides a flexible and user-friendly interface for analyzing credal partitions, along with tools for visualization and evaluation. Based on the Python ecosystem, it integrates seamlessly with popular libraries such as \texttt{numpy}, \texttt{pandas}, \texttt{matplotlib}, \texttt{scipy}, and \texttt{scikit-learn}. See Table \ref{tab:metadata} for software metadata. \\

This paper is organized as follows. Section \ref{evclustbackground} explains the core concepts behind evidential clustering. Section \ref{evclustsoft} focuses on the \evclust software, outlining its structure and main features, including clustering algorithms, utility tools, and performance metrics. Section \ref{evclustillust} demonstrates how \evclust can be used in practice. Finally, Section \ref{conclusion} discusses experimental results and wraps up with the conclusion.

\begin{table}[htbp!]
    \centering
    \caption{Metadata for the \evclust package.}
    \renewcommand{\arraystretch}{1.2}
    \begin{tabular}{l|p{9cm}}
    \hline\hline
    \textbf{Metadata} & \textbf{Description} \\ 
    \hline
    Current code version & v0.2 \\ 
    Permanent link & \url{https://github.com/armelsoubeiga/evclust} \\ 
    Legal code license & MIT License \\ 
    Code versioning system & git \\ 
    Software languages, & Python\\ 
    Compilation requirements & \texttt{Python} $>3.8$, \texttt{scipy} $>1.10$ \texttt{scikit-learn} $>1.3.0$, \texttt{numpy}, \texttt{pandas}\\ 
    Operating systems & Cross-platform (Windows, macOS, Linux) \\ 
    Documentation & Available on  \url{https://evclust.readthedocs.io} \\ 
    Support email & \texttt{armel.soubeiga@uca.fr} \\ 
    \end{tabular}
    \label{tab:metadata}
\end{table}

\section{Evidential Clustering} \label{evclustbackground}
\subsection{Theory of Belief Functions}
The theory of belief functions \cite{evclust01}, also called the Dempster-Shafer theory, has been widely applied in various fields such as classification and clustering. It provides powerful tools for modeling uncertain and imprecise information.  Let $\omega$ be a variable that takes values in a finite set $\Omega = \left \{ \omega_1, ..., \omega_c \right \}$. Partial knowledge of the actual value of $\omega$ can be represented by a mass function $m$, defined from the set $2^{\Omega}$ to $\left [ 0, 1 \right ]$ and satisfying: $\sum_{A \subseteq \Omega}^{} m(A) = 1$. The quantity $m(A)$ is then interpreted as the amount of belief allocated to any subset $A \subseteq \Omega$, and $A$ is called a focal element if $m(A) > 0$. When $m(A) = 1$, we have a certain belief. When $m(A) > 0$ and $|A| > 1$, we have an imprecise belief. 

\subsection{Evidential data and partition}

Evidential data and partition refer to the modeling of uncertainty and imprecision within the framework of belief function theory. This approach contrasts with hard or crisp data and partitions, where uncertainty and imprecision are absent, often referred to as soft in the presence of these features. In this context, datasets can be categorized into three types, included, object data (vectorial data), relational data, and functional data. Let us denote the set of $n$ objects by $\mathbf{X} = \{\mathbf{x}_1, \cdots, \mathbf{x}_n\}$, and the set of $c$ clusters by $\Omega = \{\omega_1, \cdots, \omega_c\}$. A hard partition of the dataset $\mathcal{O}$ assumes that the cluster memberships of each object are known with certainty. Such a partition can be represented by binary variables $u_{ik}$, where:
\begin{equation}
u_{ik} = 
\begin{cases} 
1 & \text{if object } \mathbf{x}_i \text{ belongs to cluster } \omega_k, \\
0 & \text{otherwise}.
\end{cases}
\end{equation}

When objects cannot be assigned to clusters with certainty, their memberships can be represented by mass functions $m^\Omega_i$ for $i = 1, \cdots, n$. Each mass $m^\Omega_i(A)$ represents the degree of belief that the true cluster of object $\mathbf{x}_i$ is in $A$, where $A \subseteq \Omega$, and no more specific proposition \cite{ev013}. Given a mass function $m$ on $\Omega$, the belief function $\text{Bel}()$ and plausibility function $\text{Pl}()$ are defined as:

\begin{equation}
\text{Bel}(A) = \sum_{B \subseteq A} m(B),
\label{eq:belief}
\end{equation}
\begin{equation}
\text{Pl}(A) = \sum_{B \cap A \neq \emptyset} m(B) = 1 - \text{Bel}(\bar{A}),
\label{eq:plausibility}
\end{equation}

where $\bar{A}$ denotes the complement of $A$ in $\Omega$. The belief function $\text{Bel}(A)$ measures the total support for the proposition $A$, while the plausibility function $\text{Pl}(A)$ quantifies the lack of support against $A$. This evidential representation provides a powerful framework for handling uncertain and imprecise data, allowing more nuanced interpretations of cluster memberships and data relationships.

\section{Description and architecture of \evclust software} \label{evclustsoft}
\subsection{Evidential Clustering Algorithms in \evclust}
Table \ref{tab:algos} presents an overview of the evidential clustering algorithms implemented in \evclust. Some of these algorithms require attribute data, whereas others can handle proximity data, i.e., a matrix of dissimilarities, or distances between objects. Each clustering algorithm integrated into \evclust includes an example section in its documentation, which provides a practical illustration of how to use the algorithm. This section serves as a guide to help users understand the implementation of the algorithm and apply it to real-world data.

\begin{table}[H]
\centering
\caption{Overview of clustering algorithms in \evclust.}
\renewcommand{\arraystretch}{1.0}
\label{tab:algos}
    \begin{tabular}{l|l|c|l}
    \hline\hline
    \textbf{Method} & \textbf{Inputs} & \textbf{Complexity} & \textbf{Function} \\ 
    \hline
    ECM & Attribute data & $\mathcal{O}(n2^c)$ & \texttt{ecm()} \\
    RECM & Proximity data & $\mathcal{O}(nc^2 + cn^2)$ & \texttt{recm()} \\
    k-EVCLUS & Proximity data & $\mathcal{O}(n^2c^2)$ & \texttt{kevclus()} \\
    CatECM & Categorical Attribute data & $\mathcal{O}(n2^c)$ & \texttt{catecm()} \\
    EGMM & Attribute data & $\mathcal{O}(n2^c)$ & \texttt{egmm()} \\
    BPEC & Attribute data & $\mathcal{O}(n^2 + n2^c)$ & \texttt{bpec()} \\
    ECMdd &  Proximity data & $\mathcal{O}(cn^2 + n2^c)$ & \texttt{ecmdd()} \\
    MECM & Attribute data & $\mathcal{O}(n2^c)$ & \texttt{mecm()} \\
    WMVEC & Multi-view Attribute data & $\mathcal{O}(np2^c)$ & \texttt{wmvec()} \\
    WMVEC-FP & Multi-view Attribute data & $\mathcal{O}(nfp2^c)$ & \texttt{wmvec\_fp()} \\
    MECMdd-RWG & Multi-view Proximity data & $\mathcal{O}(pn2^c + pcn^2)$ & \texttt{mecmdd\_rwg()} \\
    MECMdd-RWL & Multi-view Proximity data & $\mathcal{O}(pn2^c + pcn^2)$ & \texttt{mecmdd\_rwl()} \\
    CCM & Attribute data & $\mathcal{O}(n2^c)$ & \texttt{ccm()} \\
    \hline
    \multicolumn{4}{l}{\texttt{n} number of objects, \texttt{c} number of cluster}\\
    \multicolumn{4}{l}{\texttt{p} number of view, \texttt{f} number of features}\\
    \end{tabular}
\end{table}

\subsubsection{ECM -- Evidential c-Means}
Introduced by Masson and Denoeux \cite{ev00}, the evidential c-means (ECM) algorithm is an alternating optimization procedure in the same family as the hard and fuzzy c-means algorithms. In ECM, it allows the object to be in any singleton clusters, meta-clusters , and the $\emptyset$ cluster with different masses of beliefs. The objective function of ECM is defined by:

\begin{equation}
J_{ECM}= \sum_{i=1}^{n}\sum_{A_j \neq \emptyset} {\left | A_j \right |}^{\alpha} m_{ij}^{\beta} d_{ij}^{2} + \sum_{i=1}^{n} \delta^{2} m_{i\emptyset}^{\beta}.
\end{equation}

subject to 
\begin{equation}
\sum_{j / A_j \subseteq \Omega,A_j \neq \emptyset} m_{ij} + m_{i\emptyset} = 1.
\end{equation}

where $d_{ij}$ represents the Euclidean distance between the $i$-th object and the center of $j$-th cluster. The prototype (centroid) $\mathbf{\bar{v}}_j$ of meta-cluster $A_j$ is the average of the involved singleton cluster centers, defined by:

\begin{equation}
\mathbf{\bar{v}}_j = \frac{1}{|A_j|} \sum_{k=1}^{c} s_{kj} \mathbf{v}_k, \quad s_{kj} = \begin{cases} 
      1, & \text{if } \{\omega_k\} \in A_j \\ 
      0, & \text{otherwise} 
   \end{cases}
\end{equation}

where $|A_j|$ is the cardinality of $A_j$ and $\mathbf{v}_k$ is the center of the singleton cluster $\{\omega_k\}$. In \evclust, ECM is implemented in function $ecm()$.

\begin{verbatim}
ecm(x, c, g0 = NULL, type = "full", pairs = NULL, Omega = TRUE,
    ntrials = 1, alpha = 1, beta = 2, delta = 10, epsi = 0.001, disp = TRUE)
\end{verbatim}

This function requires at least two key inputs: the data matrix which has dimensions $n \times p$ ($n$ is the number of samples, and $p$ is the number of attributes), and $c$, the number of clusters. The $g0$ argument is optional and allows to provide an initial matrix of cluster prototypes. The $type$ parameter defines the focal sets:  $'full'$ ( All subsets of $\Omega$), $'simple'$ (Only the empty set, singletons, and $\Omega$), and $'pairs'$ (The empty set, singletons, $\Omega$, and either all pair subsets) or only the specific pairs provided through the $pairs$ argument. If $Omega = False$, the set $\Omega$ is excluded from the focal sets. Other parameters control aspects of the algorithm, like the number of trials ($ntrials$), the weighting coefficients ($alpha$, $beta$, $delta$), and the stopping criterion ($epsi$). One can display the progression of the algorithm by setting $disp = True$. For more details, use \texttt{help(ecm)}.

\subsubsection{RECM -- Relational Evidential c-Means}
The Relational Evidential c-Means (RECM) algorithm was introduced as an extension of the ECM algorithm to handle proximity data directly \cite{ev005}. Unlike ECM, which operates on attribute data, RECM takes as input a dissimilarity matrix $\mathbf{D} = [\tau_{ij}]$, where $\tau_{ij}$ represents the dissimilarity between objects $i$ and $j$. The objective function of RECM is defined as follows:

\begin{equation}
J_{RECM} = \sum_{i=1}^{n} \sum_{A_j \neq \emptyset} |A_j|^\alpha m_{ij}^\beta \tau_{ij} + \sum_{i=1}^n \delta^2 m_{i\emptyset}^\beta,
\end{equation}

subject to the constraint:
\begin{equation}
\sum_{A_j \subseteq \Omega, A_j \neq \emptyset} m_{ij} + m_{i\emptyset} = 1, \quad \forall i.
\end{equation}

Here, $\tau_{ij}$ is the squared distance between object $i$ and the prototype of cluster $A_j$, which is calculated using the proximity data. The optimization procedure of RECM, based on an alternate minimization scheme, is computationally much more efficient.  Although based on the assumption that the input dissimilarities are squared Euclidean distances. The complexity of RECM is $\mathcal{O}(nc^2 + cn^2)$. This algorithm is implemented in \evclust\ under the function \texttt{recm()}.  Function \texttt{recm()} has the same arguments as \texttt{ecm()}, except that the first argument is a distance matrix $\mathbf{D}$ instead of an object-attributes matrix. Also, the algorithm can be initialized with a matrix $m0$ of masses, instead of the matrix $g0$ of initial prototypes.

\subsubsection{k-EVCLUS -- k Evidential Clustering}
The k-EVCLUS algorithm, introduced by  \cite{ev011, ev010} is a proximity-based clustering method inspired by multidimensional scaling. It aims to learn a credal partition by minimizing the discrepancy between input dissimilarities and distances in a belief space. It takes as input a symmetric $n \times n$ dissimilarity matrix $\mathbf{D} = [\tau_{ij}]$, where $\tau_{ij}$ denotes the dissimilarity between objects $i$ and $j$. Dissimilarities may be computed from attribute data, or they may be directly available. The complexity of k-EVCLUS is $\mathcal{O}(n^2c^2)$, and it is implemented in \evclust\ with the function \texttt{kevclus()}.

\begin{verbatim}
kevclus(x=None, k=None, D=None, J=None, c=None, type='simple', pairs=None, m0=None, 
         ntrials=1, disp=True, maxit=20, epsi=0.001, d0=None, tr=False, 
        change_order=False, norm=1)
\end{verbatim}

\subsubsection{CatECM -- Categorical Evidential c-Means}
The Categorical Evidential c-Means (CatECM) algorithm extends ECM to handle categorical attribute data \cite{ev012}. Instead of Euclidean distances, CatECM uses a specific similarity measure adapted to categorical data (e.g. Hamming distance). The objective function retains the same form as ECM:

\begin{equation}
J_{CatECM} = \sum_{i=1}^n \sum_{A_j \neq \emptyset} |A_j|^\alpha m_{ij}^\beta d_{ij}^2 + \sum_{i=1}^n \delta^2 m_{i\emptyset}^\beta,
\end{equation}

where $d_{ij}$ is computed based on a categorical dissimilarity measure. The algorithm has a complexity of $\mathcal{O}(n2^c)$ and is implemented in \evclust\ as \texttt{catecm()}.

\subsubsection{EGMM -- Evidential Gaussian Mixture Model}
The Evidential Gaussian Mixture Model (EGMM) \cite{ev015} is a probabilistic approach that combines Gaussian mixture using and covariance matrices of the Gaussian components. The complexity is $\mathcal{O}(n2^c)$, and the function is \texttt{egmm()}.

\subsubsection{BPEC -- Belief Peak Evidential Clustering}
The Belief Peak Evidential Clustering (BPEC) algorithm \cite{ev016} works in a way similar to ECM, with the main difference being how the cluster centers are chosen. Instead of initializing the centers randomly, they are determined as belief peaks on a two-dimensional $\delta$-Bel graph. These peaks represent points with a high density, estimated using a K-nearest neighbors method, and are far from other points with even higher densities. Once the belief peaks are identified, the ECM algorithm is applied while keeping these centers fixed. \\

In the \evclust library, the \texttt{delta\_Bel()} function is used to generate the $\delta$-Bel graph. It takes as input the data matrix, the number of neighbors \texttt{K}, and a scale parameter \texttt{q}. This function plots the $\delta$-Bel graph, and the user is prompted to select the bottom-right corner of a rectangle enclosing the belief peaks. The identified peaks are returned as output. After determining the belief peaks, the \texttt{bpec()} function can be executed. Its syntax is nearly identical to that of \texttt{ecm()}, but with the belief peaks set as fixed cluster centers. The computational complexity of BPEC is $\mathcal{O}(n^2 + n2^c)$.

\subsubsection{ECMdd -- Evidential c-Medoids}
ECMdd \cite{ev004} extends ECM for clustering based on medoids rather than centroids, which makes it robust to noise and outliers. It takes as input a symmetric $n \times n$ dissimilarity
matrix $\mathbf{D} = [\tau_{ij}]$, where $\tau_{ij}$ denotes the dissimilarity between objects $i$ and $j$. Dissimilarities may be computed from attribute data, or they may be directly available. They need not
satisfy the axioms of a distance such as the triangular inequality. The objective function is:

\begin{equation}
J_{ECMdd} = \sum_{i=1}^n \sum_{A_j \neq \emptyset} |A_j|^\alpha m_{ij}^\beta \tau_{ij}^2 + \sum_{i=1}^n \delta^2 m_{i\emptyset}^\beta.
\end{equation}

where medoids are selected as representative objects for clusters. Its complexity is $\mathcal{O}(cn^2 + n2^c)$, and it is implemented in \texttt{ecmdd()}.

\subsubsection{MECM -- Median Evidential c-Means}
The Median Evidential c-Means (MECM) \cite{ev017} is a extend version of ECM for partitioning relational data designed to handle noise and outliers by modifying the distance computation. The median variant relaxes the restriction of a metric space embedding for the objects but constrains the prototypes to be in the original data set. The theoretical definition and development in \evclust is similar to ECMdd,  see details \cite{ev017}. The complexity is $\mathcal{O}(n2^c)$, and it is implemented as \texttt{mecm()}.

\subsubsection{WMVEC -- Weighted Multi-View Evidential Clustering}
The WMVEC algorithm \cite{ev007} and its variant Adaptive weighted multi-view evidential clustering with feature preference (WMVEC-FP) \cite{ev007}, based on multi-view Attribute data,  generalizes evidential clustering to multi-view data by assigning weights to views based on their relevance and also compute feature preference. WMVEC computes the relevance of views only, while WMVEC-FP computes the relevance of both views and features. The implementations are respectively \texttt{wmvec()} and \texttt{wmvec\_pf()}. 

\subsubsection{MECMdd -- Multi-View Evidential C-Medoids}
Multi-View Evidential C-Medoid clustering with adaptive weightings (MECMdd) \cite{ev019} is similar to WMVEC but based on relational data. Based to ECMdd, MECMdd has four variants based on different weight optimization schemes. The main objective function is defined by \ref{J-MECMdd-RWL} and with a time complexity of $\mathcal{O}(pn2^c + pcn^2)$. It takes as input a symmetric $n \times n$ dissimilarity matrix $\mathbf{D} = [\tau_{ij}]$, where $\tau_{ij}$ denotes the dissimilarity between objects $i$ and $j$. Dissimilarities may be computed from attribute data, or they may be directly available. They need not satisfy the axioms of a distance such as the triangular inequality. In evclust, functions \texttt{mecmdd\_rwl()} and \texttt{mecmdd\_rwg()} are used respectively to learn MECMdd-RWL and MECMdd-RWG \cite{ev019}.

\begin{equation}\label{J-MECMdd-RWL}
J_{MECMdd} = \sum_{i=1}^{n}\sum_{A_j \neq \emptyset} {\left | A_j \right |}^{\alpha} m_{ij}^{\beta} \sum_{l=1}^{p}\left ( \lambda_{jl} \right )^s \tau_{ijl}
 + \sum_{l=1}^{p} \delta_l^{2} \sum_{i=1}^{n} m_{i\emptyset}^{\beta};
\end{equation}

\subsubsection{CCM -- Credal C-Means}
The Credal C-Means (CCM) algorithm \cite{ev018} extends ECM, involving a redefinition of the distance between the object and the centers. In CCM, the distance $D_{ij}$ between the $i^{th}$ object and the meta-cluster $A_j$ is not only related to the center $\mathbf{\bar{v}}_j$ of $A_j$ but also related to the center $\mathbf{v}_k$ of associated singleton cluster included in $A_j$ defined by :

\begin{equation}
     D_{ij} =  \frac{\sum_{\left\{w_k \right\} \in A_j} d_{ik}^2 + \gamma d_{ij}^2}{\left| A_j\right| +\gamma}
\end{equation}

where $\gamma$ represents the weighting factor for the distance from the object to the centers of meta-clusters, commonly employed to regulate the number of objects in meta-clusters. CCM is particularly useful for datasets with high uncertainty or noise, as it provides a more nuanced representation of cluster assignments by accounting for imprecision and conflict. CCM is implemented in \evclust under the function \texttt{ccm()}, which has a syntax similar to \texttt{ecm()} with $\mathcal{O}(n2^c)$ time complexity.

\subsection{Utils Functions in \evclust}
The \evclust Python library provides a set of powerful utility functions, see Table \ref{tab:utils}, designed to facilitate the exploration and analysis of credal partitions produced by evidential clustering algorithms. These functions allow users to summarize, transform, and visualize credal partitions, making them highly versatile for clustering tasks in various domains. \\

Credal partitions, which are more general than traditional clustering outputs, can be converted into classical structures such as hard, fuzzy, or rough partitions \cite{ev008}. This transformation is achieved using functions like \texttt{extractMass()}, which computes key outputs such as the plausibility and belief distributions, lower and upper approximations of clusters, and other derived metrics. Visualization is another critical aspect supported by \evclust. The library includes functions such as \texttt{ev\_plot()} for representing clusters and approximations in a two-dimensional space, \texttt{ev\_pcaplot()} for PCA-based cluster visualization, and \texttt{ev\_tsplot()} for analyzing clustering results in time-series data. These tools provide an intuitive understanding of the results and help to interpret complex clustering structures. In addition to these core functionalities, \evclust offers tools for creating and managing focal sets (\texttt{makeF()}), summarizing the properties of credal partitions (\texttt{ev\_summary()}), and detecting potential outliers. These utilities not only enhance the usability of evidential clustering but also enable a deeper exploration of uncertainty and ambiguity in cluster memberships, making \evclust a valuable resource for data scientists and researchers. Table~\ref{tab:utils} summarizes the main utility functions available in \evclust. \\

The \evclust library provides access to several real-world example datasets through its \texttt{datasets} module. These datasets, such as the Decathlon dataset from \texttt{load\_decathlon()}, the well-known Iris dataset \texttt{load\_iris()}, Protein data \texttt{load\_protein()}, the fourclass dataset \texttt{load\_fourclass()}, and the multi-view ProPt dataset \texttt{load\_prop()}, are available for users to explore and apply clustering algorithms.

\begin{table}[H]
\centering
\caption{Utility functions in \evclust.}
\renewcommand{\arraystretch}{1.2}
\label{tab:utils}
\begin{tabular}{l|p{10cm}}
\hline\hline
\textbf{Function} & \textbf{Description} \\ 
\hline
\texttt{makeF()} & Creation of a matrix of focal sets. \\ 
\texttt{ev\_summary()} & Extracts basic information to summarize a credal partition. \\ 
\texttt{ev\_plot()} & Generates plots of a credal partition. \\ 
\texttt{ev\_pcaplot()} & Plots PCA results with cluster colors of a credal partition. \\ 
\texttt{extractMass()} & Computes different outputs from a credal partition. \\ 
\end{tabular}
\end{table}

\subsection{Metrics Functions in \evclust}
The \evclust library includes a set of metrics functions that facilitate the evaluation and comparison of credal partitions. These functions address key aspects such as uncertainty quantification, relational representations, and the comparison of clustering results, offering versatile tools for evidential clustering analysis. \\

One of the core metrics implemented in \evclust is the measure of \textit{nonspecificity()}, which quantifies the imprecision of a credal partition. Introduced by Klir and Wierman \cite{ev009}, Non-Specificity accounts for the uncertainty inherent in mass functions, distinguishing between imprecision and conflict. This metric is particularly useful for analyzing the degree of uncertainty in credal partitions and has been employed to determine the number of clusters in evidential clustering \cite{ev00}. To facilitate the comparison of credal partitions, \evclust implements the \texttt{credalRI()} function, which computes the Rand Index (RI) adapted for evidential clustering. This function compares two credal partitions by first transforming them into their relational representations and then evaluating the level of agreement between them. The relational representation of a credal partition, computed using \texttt{pairwise\_mass()}, is expressed as a set of three relational matrices: $M_e$, $M_1$, and $M_0$, corresponding to the masses assigned to the empty set, singleton sets, and their complements, respectively. The \texttt{credalRI()} function then uses these representations to calculate the RI while allowing the user to specify a weighting scheme via its \texttt{type} argument. Overall, these metrics functions provide robust tools for evaluating evidential clustering results, enabling users to assess the level of uncertainty, extract insights from relational structures, and compare multiple credal partitions effectively.

\begin{table}[H]
\centering
\caption{Metrics functions in \evclust.}
\renewcommand{\arraystretch}{1.2}
\label{tab:metrics}
\begin{tabular}{l|p{10cm}}
\hline\hline
\textbf{Function} & \textbf{Description} \\ 
\hline
\texttt{nonspecificity()} & Computes the non-specificity of a credal partition. \\ 
\texttt{credalRI()} & Computes the Rand Index to compare two credal partitions. \\ 
\texttt{pairwise\_mass()} & Computes relational representations for credal partition.\\
\end{tabular}
\end{table}

\section{Illustrations}\label{evclustillust}
In this section, we demonstrate the use of the main functions in the \evclust library through
the analysis of some datasets.

\subsection{Example with Iris dataset}
This section demonstrates the use of \evclust for evidential clustering with the Iris dataset. The Iris dataset includes measurements of sepal length, sepal width, petal length, and petal width for 50 flowers from each of three iris species. Assuming uncertainty in species assignments and the possibility of overlapping species across clusters, we employ the \texttt{ecm} algorithm to analyze the data and account for these uncertainties.

\begin{verbatim}
    # Import requirements
    import evclust
    from evclust.ecm import ecm
    from evclust.datasets import load_iris

    # Import test data
    df = load_iris()
    df = df.drop(['species'], axis = 1) # del label column
\end{verbatim}

The \texttt{ecm} algorithm can be applied to these data by running function \texttt{ecm()}:

\begin{verbatim}
    # ECM with c=3
    clus = ecm(x=df, c=3, beta = 2,  alpha=1, delta=10)
\end{verbatim}

We can summary the output of the ecm model, to see Focal sets or Number of outliers

\begin{verbatim}
    from evclust.utils import ev_summary
    ev_summary(clus)
    
    ------ Credal partition ------
    3 classes,
    150 objects
    Generated by ecm
    Focal sets:
    [[0. 0. 0.]
     [1. 0. 0.]
     [0. 1. 0.]
     [1. 1. 0.]
     [0. 0. 1.]
     [1. 0. 1.]
     [0. 1. 1.]
     [1. 1. 1.]]
    Value of the criterion = 38.82
    Nonspecificity = 0.22
    Prototypes:
    [[7.06131634 3.03675091 6.05972886 2.1474559 ]
     [4.96375502 3.3462016  1.49213248 0.24695422]
     [6.01335287 2.76720722 4.77762377 1.64225065]]
    Number of outliers = 0.00
\end{verbatim}

\begin{figure}[H]
\centerline{\includegraphics[width=1\textwidth]{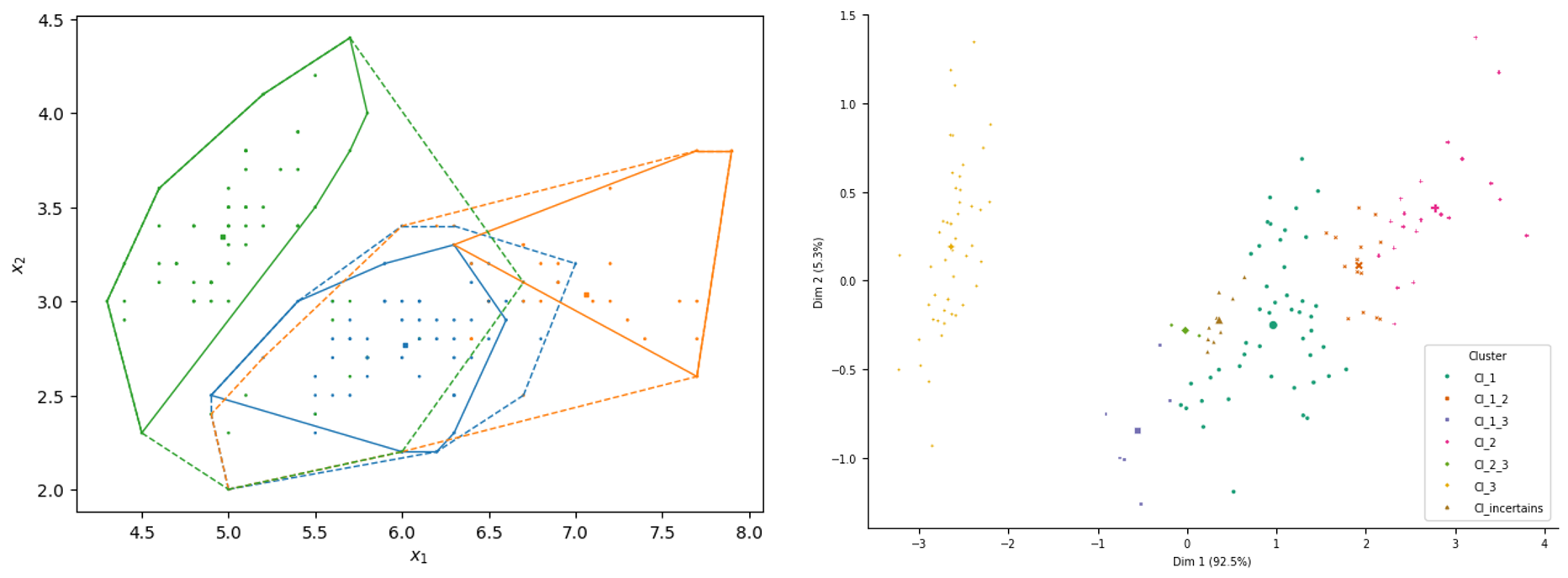}}
\caption{Illustration of visualization of credal partition using \texttt{ev\_plot} and \texttt{ev\_pcaplot} functions.}
\label{fig:evclustplot}
\end{figure}

The output of the \texttt{ev\_summary} function provides an overview of the credal partition obtained from the \texttt{ecm} clustering algorithm. In this case, three clusters were identified across 150 objects. The focal sets represent the subsets of clusters considered during the evidential clustering process, including single clusters and combinations of clusters. The prototypes indicate the centroids of the identified clusters in the feature space, showing the average characteristics of each cluster. The clustering criterion achieved a value of \texttt{Value of the criterion = 38.82}, indicating the fit of the model to the data, while the \texttt{Nonspecificity = 0.22} reflects the degree of imprecision in the partitioning. No outliers were detected in this analysis, suggesting a robust clustering outcome. \\
Figure \ref{fig:evclustplot} shows a simple plot on the left and a plot on the axes of a PCA on the right.

\begin{verbatim}
    from evclust.utils import  ev_plot, ev_pcaplot

    ev_plot(x=clus, X=df)
    ev_pcaplot(data= df, x=clus, normalize=False, splite=False, cex=8, cex_protos=5)
\end{verbatim}

\subsection{Example with multi-view dataset}
We illustrate the example of multi-view evidential clustering with the WMVEC algorithm using the \texttt{wmvec} function from \evclust.  The function is configured to learn $c=4$ clusters and for \texttt{type=“simple”} means that we are only interested in the focal set equal to 2.
\begin{verbatim}
    #Import of modules
    from evclust.wmvec import wmvec
    from evclust.datasets import load_prop

    # Multi-view data import
    df = load_prop()

    # wmvec clustering fit
    clus = wmvec(X=df, c=4, alpha=2, delta=5, maxit=20, epsi=1e-3,
                beta=1.1, lmbda=403, type="simple", disp=True)
\end{verbatim}

The Prop dataset contains three attribute views with different dimensions. When imported using the \texttt{load\_prop()} function, it returns a list containing these three views, each represented as a DataFrame. The importance weight of each view can be accessed through \texttt{clus['param']['R']}. This allows users to analyze the different attribute views and their corresponding weights in clustering tasks.

\begin{verbatim}
    clus['param']['R']

    [0.31174353, 0.34414724, 0.34410923]
\end{verbatim}

\section{Conclusion and Perspectives}\label{conclusion}
Evidential clustering offers a novel framework for clustering, where uncertainty about cluster memberships is modeled using Dempster-Shafer mass functions. This approach provides a richer representation of imprecision and ambiguity compared to traditional clustering methods, making it especially suitable for complex and uncertain datasets. \\

In this paper, we introduced \evclust, a Python library that implements a comprehensive set of evidential clustering algorithms. These algorithms are designed to address a wide range of clustering challenges, including handling non metric dissimilarity data, discovering clusters with complex shapes, and performing model-base clustering. Additionally, \evclust provides tools for visualizing, evaluating, and exploiting credal partitions. The open-source nature of the library and its implementation in Python ensure broad accessibility and seamless integration into various workflows, benefiting from the widespread use of Python in the data science community. \evclust caters to a diverse audience, including researchers, data analysts, students, programmers, and decision-makers, with applications extending beyond clustering to data analysis and decision-making under uncertainty. \\

We plan to expand \evclust in several directions. Our road-map includes adding new algorithms to address emerging challenges in data science, such as Belief Shift Clustering (BSC), Dynamic evidential c-means clustering (DECM), Deep Evidential Clustering (DEC), Decision tree-based evidential clustering (DTEC), Transfer learning-based evidential c-means clustering (TECM), etc. We also aim to enhance the library's usability through simplified documentation and tutorial notebooks, making it more accessible for educational and practical purposes. Encouraging community contributions via platforms like GitHub will ensure continuous improvement and adaptation to the evolving needs of users. We aim to improve usability with simplified documentation and tutorials, fostering accessibility for education and practice, while encouraging community contributions on GitHub for ongoing development. \\

In conclusion, \evclust is a powerful tool for evidence clustering and beyond, with significant potential for growth and impact in the data science community.

\section*{Acknowledgments}
The authors acknowledge the support received from the Agence Nationale de la Recherche of the French government through the program 16-IDEX-0001 CAP 20-25.

\bibliographystyle{unsrt}
\bibliography{bibliography}

\end{document}